\documentstyle[emulateapj,times,psfig]{article}

\def\beq{\begin{equation}} \def\eeq{\end{equation}} 
\def\microJy{{\rm \mu Jy}} \def\cgs{{\rm erg\, cm^{-2}\, s^{-1}}}
 \def\eg{e.g$.$~}   \def\etal{et al$.$~}
\def\epsel{\varepsilon_e} \def\epsmag{\varepsilon_B} \def\cm3{\;{\rm cm^{-3}}}
\def\simg{\mathrel{%
      \rlap{\raise 0.511ex \hbox{$>$}}{\lower 0.511ex \hbox{$\sim$}}}}
\def\siml{\mathrel{%
      \rlap{\raise 0.511ex \hbox{$<$}}{\lower 0.511ex \hbox{$\sim$}}}}

\begin{document}

\title{Observational Prospects for Afterglows of Short Duration Gamma-ray Bursts}

\author{A. Panaitescu} 
\affil{Dept. of Astrophysical Sciences, Princeton University,
Princeton, NJ 08544} 
\author{P. Kumar} 
\affil{Institute for Advanced Study, Olden Lane, Princeton, NJ 08540}
\author{R. Narayan} 
\affil{Harvard-Smithsonian Center for Astrophysics, Cambridge, MA
02138}

\begin{abstract}

 If the efficiency for producing $\gamma$-rays is the same in short
duration ($\siml 2$ s) Gamma-Ray Bursts (GRBs) as in long duration
GRBs, then the average kinetic energy of short GRBs must be $\sim 20$
times less than that of long GRBs.  Assuming further that the
relativistic shocks in short and long duration GRBs have similar
parameters, we show that the afterglows of short GRBs will be on
average 10--40 times dimmer than those of long GRBs.  We find that the
afterglow of a typical short GRB will be below the detection limit
($\siml 10\,\microJy$) of searches at radio frequencies.  The
afterglow would be difficult to observe also in the optical, where we
predict $R \simg 23$ a few hours after the burst.  The radio and
optical afterglow would be even fainter if short GRBs occur in a
low-density medium, as expected in NS--NS and NS--BH merger models.
The best prospects for detecting short-GRB afterglows are with early
($\siml 1$ day) observations in X-rays.

\end{abstract}

\keywords{gamma-rays: bursts - ISM: jets and outflows - radiation
mechanisms: non-thermal}

\section{Introduction}

 Kouveliotou \etal (1993) showed that the durations of Gamma Ray
Bursts (GRBs) have a bimodal distribution, with roughly a third of the
observed bursts corresponding to a short duration ($\siml2$ s)
population, and the remaining two-thirds belonging to a long duration
($\simg2$ s) population.

 Searches for transient X-ray, optical and radio afterglow emission
have so far been largely limited to long duration bursts.  This is
because the Italian-Dutch BeppoSAX satellite and other presently
active instruments, which have been used to obtain well-determined GRB
coordinates, are sensitive only to bursts longer than a few seconds.
Most of the X-ray follow-ups of long GRBs have been successful, and 35
X-ray afterglows have been detected so far (mid-2001). The success
rate of optical searches has been somewhat less (Lazzati,
Covino \& Ghisellini 2001; Fynbo \etal 2001), and 23 optical
afterglows have been detected; the majority of optical afterglows have
yielded redshifts.  The success rate in radio has been comparable to
that in the optical. 

Among short GRBs, optical and radio searches for afterglows have so
far been carried out for only four sources that happened to be well
localized by the Interplanetary Network (Hurley \etal 2001).  No
afterglow emission was detected, but the sensitivity of the searches
was not very high.  More searches will be carried out in the future
when rapid arc-minute localizations of short GRBs become routinely
available from the upcoming HETE II and Swift satellites.

In this {\it Letter}, we make reasonable assumptions about the
physical parameters of short GRBs and estimate the broadband afterglow
emission to be expected from these bursts.  The results may be useful
for designing search strategies for afterglows of short GRBs.

\section{Short versus Long GRBs}

 The predictions for the light-curves of the afterglows depend on some
key parameters.  We attempt to estimate the relative magnitudes of
these parameters in short and long GRBs using observational data and
some theoretical ideas.

Studies of the temporal and spectral properties of short and long
GRBs have revealed that the burst duration is anticorrelated with the
spectral hardness (Kouveliotou \etal 1993).
The peak frequency of the $\nu F_\nu$ spectrum increases with the bulk
Lorentz factor of the GRB in the external shock model for producing
$\gamma$-ray emission (Rees \& M\'esz\'aros 1992; Piran, Shemi \& Narayan 
1993; Katz 1994), however the relationship is complicated in the internal shock
model (see Piran 1999 for a review). Fortunately, the
Lorentz factor (LF) of the late-time GRB remnant which produces the
afterglow is practically independent of the LF during the
early GRB phase, and so it is unlikely that any difference in the 
initial LF for the long and the short duration GRBs will have any
effect on the afterglow flux. We note, however, that if the initial jet opening 
angle $\theta_j$ for short duration GRBs were larger than
longer lasting GRBs, it will cause the afterglows of short
GRBs to become considerably dimmer since the jet transition time (see
\S3) varies as $t_j\propto\theta_j^{8/3}$.

 By analysing a sample of over 400 GRBs, and allowing for
observational bias against detecting weak and long bursts with peak
fluxes below the detection threshold, Lee \& Petrosian (1997) showed
that there is a highly significant positive correlation between the
burst fluence and duration.  In another study, Mukherjee \etal (1998)
used two multivariate clustering methods to show that most of the
structure in the multidimensional space of burst observations is
contained in three fundamental quantities: duration, fluence, and
spectral hardness.  According to their analysis, short/hard GRBs have
a 25 keV--1 MeV fluence $\Phi$ (in cgs units) of approximately $\log
\Phi = -6.4 \pm 0.6$, while long/soft GRBs have $\log \Phi = -5.2 \pm
0.6$.  Short bursts thus have about 20 times less fluence than long
bursts (see also Mao et al. 1994, Piran 1996).

 Figure 1 shows the 25 keV--1MeV fluences $\Phi$ and the durations
$T_{90}$ of 34 long GRBs for which
afterglows have been observed.  The average duration for these GRBs is
$\log T_{90} = 1.5 \pm 0.5$, similar to that of the 486 long bursts
analyzed by Mukherjee \etal (1998) from the Third BATSE Catalog
(Meegan \etal 1996). The average fluence of long GRBs with afterglows
is $\log \Phi = -4.8\pm0.72$, implying that these bursts are, on
average, approximately twice as bright as the long bursts analyzed by
Mukherjee et al. (1998) and 40 times more energetic than the 203 short
bursts analyzed by them.  

 The GRB energy output is determined by the kinetic energy of the
relativistic outflow and the efficiency with which dissipative and
radiative mechanisms convert some of the energy into $\gamma$-ray
emission. Assuming that GRBs arise from internal shocks in unsteady
winds, the dissipation efficiency is determined by the magnitude of
the fluctuations in the ejection Lorentz factors of various parts of
the outflow. Our present understanding of the properties of GRB
progenitors does not allow us to establish a correlation between the
GRB duration and efficiency.  It is possible that short GRBs are
somewhat less efficient than long GRBs because the radius at which
internal shocks occur in short GRBs is closer to the photospheric
radius (Kumar 1999), so that some of the emission may be degraded by
multiple scattering. Ignoring this effect, we expect the ratio of the 
kinetic energies of short and long GRBs to be the ratio of their
fluences, i.e., a factor of about 20.

 We finally discuss the environments in which short and long bursts
take place.  The existence of these two clearly distinct populations
of GRBs probably implies different physical origins.  Most models of
GRBs -- mergers of binary neutron stars (NS--NS), black hole (BH)--NS,
or BH--white dwarf (WD), and failed supernovae/collapsars -- involve
the formation of a BH surrounded by a disk of debris.  The spin energy
of the hole and the gravitational, thermal, and rotational energy of
the disk represent the available reservoirs to power the $\simg
10^{51}$ ergs required for the GRB .  The BH spin may be tapped by the
Blandford-Znajek process (1977), and the disk energy may be extracted
either via neutrino annihilation (Eichler et al. 1989) or via magnetic
fields/flares (Narayan, Paczy\'nski \& Piran 1992; see \eg
M\'esz\'aros, Rees \& Wijers 1999 for a review).

 The burst duration is determined both by the timescale for ejecting
the relativistic outflow, and by the processes that $i)$ shape the
outflow dynamics until the gas reaches the region where the
$\gamma$-ray photons are emitted, such as the penetration of the GRB
jet through the envelope of a collapsed star, and $ii)$ convert the
outflow kinetic energy into $\gamma$-rays. In those bursts where the
high energy emission arises as the outflow energy is dissipated
through interaction with the circumburst medium, GRBs of long duration
can be obtained even if the initial ejection is impulsive. However,
the general absence of signatures of an external shock in the temporal
structure of GRBs (Sari \& Piran 1997; Ramirez-Ruiz \& Fenimore 2000)
indicates that the GRB emission is produced in internal shocks over a
small range of distances.  If this is the case, then the GRB duration
is a direct measure of the time interval during which the "central
engine" is active.

 If GRBs originate in NS--NS or NS-BH mergers (Goodman 1986; Eichler
\etal 1989; Paczy\'nski 1991; Narayan \etal 1992; M\'esz\'aros \& Rees
1992; Katz \& Canel 1996), enough mass reaches the resulting BH to
power the GRB, provided the disk is sufficiently small and accretion
is driven by neutrino cooling (Narayan, Piran \& Kumar 2001).  The
expected duration of the relativistic wind (and thus the GRB) is under
1 s.  Thus these progenitors can naturally produce short GRBs.

 The collapse of the iron core of a massive star (collapsar model:
Woosley 1993; Paczy\'nski 1998) with intermediate angular momentum
leads to the formation of a disk with a sufficient accretion rate to
power a relativistic outflow lasting for 10--20 s (MacFadyen \&
Woosley 1999).  The collapsar model is thus more appropriate for long
GRBs. If the core collapse produces a NS and an out-going shock, the
BH and the torus form when the ejected matter, lacking sufficient
momentum, falls back (MacFadyen, Woosley \& Heger 2001). In this model
the GRB duration is set by the dynamics of the fallback, and most
likely would accommodate only the longest GRBs, lasting for more than
100 s.  Long GRBs could also arise from WD--BH (Fryer \etal 1999) or
helium star--BH mergers (Zhang \& Fryer 2001), which lead to the
formation of larger disks with accretion times above 10 s. However, it
is likely that the gas will accrete via a convection-dominated
accretion flow in these systems, in which case the flow is likely to
be rather inefficient at extracting the disk energy (Narayan \etal
2001).
 
 According to these ideas, then, there is a clear association between
the type of GRB progenitor and the burst duration.  Short GRBs are
likely to involve merging NS binaries (NS--NS or NS--BH), which should
occur predominantly in the low density halo of the host galaxy, given
the velocities of a few hundred ${\rm km\,s^{-1}}$ acquired by the NS
at birth and the binary coalescence time of about 100 Myr.
Long GRBs, on the other hand, are likely to be associated with massive
stars that at the end of their evolution (a few Myr) are still within
the cloud where they formed.  We note that Bloom, Kulkarni \&
Djorgovski (2001) have identified the host galaxies of 20 afterglows
of long duration GRBs and have found GRB--host offsets lower than
those expected for merging NS binaries (Bloom, Sigurdsson \& Pols
1999; Bulik, Belczy\'nski \& Zbijewski 1999; Fryer, Woosley \&
Hartmann 1999).  Furthermore, the high column densities identified by
Owens \etal (1998) and Galama \& Wijers (2001) from the absorption of
the X-ray afterglow emission indicate that long-duration GRBs occur in
giant molecular clouds, yielding further support for the massive star
collapse model for long bursts.

 If the above arguments are correct, we expect the afterglows of short
GRBs to occur in a more tenuous medium compared to the afterglows of
long-duration GRBs.  The difference in the density of the medium is
likely to be a few orders of magnitude.

\section{Afterglows of Short GRBs}

 Given the kinetic energy $E$ of the relativistic outflow and the
number density $n$ of the ambient external medium, plus a few other
dimensionless parameters described below, the afterglow light-curves
in various bands may be theoretically estimated (e.g., Sari, Piran \&
Narayan 1998; Panaitescu \& Kumar 2001a, and references therein).  If
we consider early times, when the radio emission is below the peak
frequency $\nu_p$ of the afterglow spectrum, and if we assume that the
cooling frequency $\nu_c$ is between the optical and X-ray bands, then
the analysis of Panaitescu \& Kumar (2001a) shows that 
\beq 
F_{radio}
\propto \epsel^{-2/3} \epsmag^{1/3} E^{5/6} n^{1/2} \;,
\label{radio}
\eeq
\beq
 F_{optical} \propto \epsel^{p-1} \epsmag^{(p+1)/4} E^{(p+3)/4} n^{1/2} \;,
\label{optic}
\eeq
\beq
 F_{X-ray} \propto \epsel^{p-1} \epsmag^{(p-2)/4} E^{(p+2)/4} \;.
\label{xray}
\eeq Here, $E$ is the isotropic equivalent kinetic energy of the
outflow, $\epsel$ is the fractional energy in electrons, $p$ is the
index of the power-law distribution of the electron Lorentz factor
$N(\gamma) \propto \gamma^{-p}$, and $\epsmag$ is the fraction of the
post-shock thermal energy in magnetic field (see Panaitescu \& Kumar
2001a for details).  The break frequencies satisfy $\nu_p \propto
E^{1/2} \epsel^2 \epsmag^{1/2}$, $\nu_c \propto E^{-1/2} n^{-1}
\epsmag^{-3/2}$ for synchrotron-dominated electron cooling, and $\nu_c
\propto [E^{-p/2} n^{-2} \epsel^{-2(p-1)} \epsmag^{-p/2}]^{1/(4-p)}$
when inverse Compton losses exceed synchrotron cooling.

 The above equations show that for $p \sim 2$, which is the average
value for this parameter determined by Panaitescu \& Kumar (2001b)
from numerical modeling of eight GRB afterglows, the afterglow flux is
roughly proportional to the kinetic energy $E$.  Since short GRBs have
a fluence on average $\sim 20$ times smaller than that of long GRBs,
other parameters being equal, we expect the afterglows of short GRBs
to be $\sim 20$ times dimmer than the afterglows of long GRBs.

 The equations show that the afterglow flux in short bursts would be
further suppressed if these bursts occur in a lower density medium
compared to long bursts.  The expressions for the radio and optical
flux have an explicit dependence on $n$.  Even though the formula for
the X-ray flux does not have an explicit dependence, this flux is also
suppressed since, for low $n$, the cooling frequency $\nu_c$ moves
above the X-ray band and so the X-ray afterglow light-curve is
described by equation (\ref{optic}) rather than equation (\ref{xray}).

 Finally, if the breaks observed at one to a few days in the optical
emission of several GRB afterglows are due to collimation of the
outflow (Rhoads 1999; Sari et al. 1999; Kumar \& Panaitescu 2000), 
then the $\sim 20$ times smaller
kinetic energy of the jets in short GRBs will cause the jet break time
$t_j \propto E^{1/3}$ to occur $\sim 3$ times earlier in these bursts
compared to long GRBs. This will further diminish the brightness of
short-GRB afterglows (at a fixed observing time) by a factor of about
$2-3$.

 These conclusions are illustrated in Figure 2, which compares the
radio, optical, and X-ray emission of a typical long-GRB afterglow
with that predicted for short-GRB afterglows with lower values of $E$
and $n$.  The numerical calculations of the afterglow dynamics and
emission of radiation were done by the methods described in Panaitescu
\& Kumar (2000). The parameters for the long-GRB afterglow were chosen
such that at one day after the burst the model yields a radio flux of
$\sim 0.5$ mJy, an optical brightness of $R \sim 20$ and a 2--10 keV
X-ray flux of $\sim 10^{-11} \cgs$, in rough agreement with
observations.

 We see from Fig. 2 that the radio emission from a typical short-GRB
afterglow will most likely be very difficult to detect. The same is
true also for the optical emission, where only early ($\siml 1/2$ day)
observations as deep as the measurements made by Groot \etal (1998)
for GRB 970828 and by Fynbo \etal (2001) for GRB 000630 are likely to
be successful.  The best chance for detecting the afterglow of a short
GRB is with X-ray observations.  BeppoSax will need to observe earlier
than $\sim 1/2$ day, and CXO, HETE II and Swift earlier than $\sim 2$
days.  Such observations are quite feasible.
 
 Because of the roughly linear dependence of the afterglow brightness
on the kinetic energy $E$, any difference in the average redshifts of
short and long GRBs will not alter the above
conclusions.  The redshift enters into the estimate of $E$, and
through it in the estimate of the afterglow luminosity, but it factors
out when the luminosity is converted to the observed flux.

\section{Conclusions}

 The fluences above 25 keV of GRBs detected by BATSE are positively
correlated with the durations of these bursts; long bursts (durations
longer than a few seconds) are on average 20 times more energetic than
short bursts.  Moreover, the $\sim$ three dozen long GRBs for which
afterglows have been detected so far (following their accurate
localization by BeppoSAX and the Interplanetary Network) are on
average 40 times brighter (in fluence) than typical short GRBs.  Thus
we expect the relativistic outflow of a long GRB to have about 20
times the kinetic energy of the outflow from a short GRB. If
the other parameters that influence the dynamics of the GRB remnant
and the emission of radiation are roughly the same, we predict that
the afterglows of short GRBs should be a factor of 10 or more
dimmer than the afterglows of long GRBs.

 If short GRBs arise from merging NS--NS or NS--BH binaries which may
have traveled upto or beyond the very tenuous outskirts of their host
galaxies, and if long GRBs are due to the core collapse of massive
stars which die in the dense molecular clouds where they were formed,
then the densities $n$ of the media surrounding these two types of
GRBs may differ by a few orders of magnitude.  This could further
seriously diminish the prospects of detecting radio and optical
afterglows of short GRBs since the afterglow brightness in these bands
is proportional to $n^{1/2}$.

 We conclude that the best chance of detecting afterglows of short
GRBs is with early X-ray observations within 1 day after the
GRB. Rapid and deep optical follow-up within a few hours after the
main event may also lead to a detection.  Radio observations appear
the least promising (Fig. 2).  

Of course, short GRBs significantly brighter than average could be as
energetic as some of the long GRBs for which afterglows have been seen
(Fig. 1). The afterglows of such unusually bright short GRBs could be
detected even beyond a day, particularly if the magnetic field
strength is close to equipartition ($\epsmag\to1$).  But such bursts
should be in the minority.  On the other hand, if future observations
indicate that the afterglows of most short GRBs are as bright as those
of long GRBs, it would imply that one or more of the assumptions we
have made in our analysis is invalid.  One possibility is that the
$\gamma$-efficiency is significantly lower in short bursts than in
long bursts, as disussed in Kumar (1999), and another is that the
efficiency spans a wide range in both types of bursts, indicating a
highly inhomogeneous outflow (Kumar \& Piran 2000).

\acknowledgments{AP acknowledges support from Princeton University
                 through the Lyman Spitzer, Jr. fellowship. PK was
                 supported in part by the Ambros Monell Foundation and
                 NSF grant PHY-0070928, and RN was supported in part
                 by NSF grant AST-9820686. We are grateful to Tsvi Piran
                 for comments.}

\begin{figure*}
\centerline{\psfig{figure=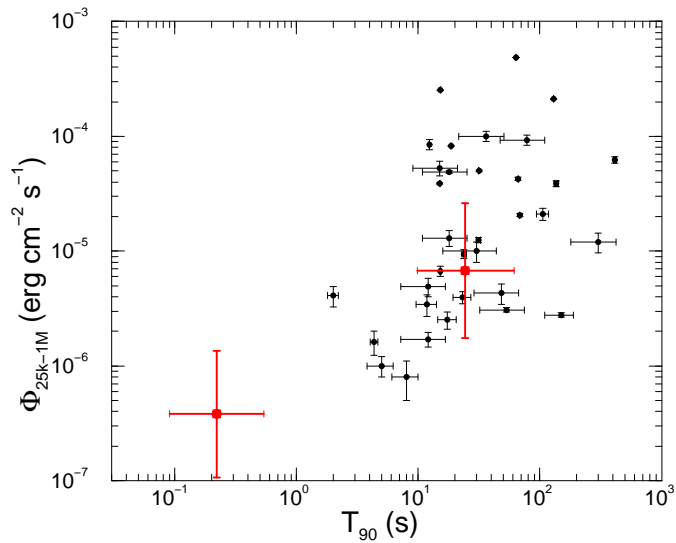}} \figcaption{ 25 keV--1 MeV
fluence $\Phi$ versus duration $T_{90}$ for 34 long GRBs for which a
transient emission (afterglow) has been observed.  Data for 21 GRBs
are taken from the current BATSE catalog
(http://www.batse.msfc.nasa.gov/batse/), and the rest are from the GCN
Archive (http://gcn.gsfc.nasa.gov/gcn/).  For the latter, a 10\%-20\%
uncertainty in fluence was assumed and the quoted burst durations were
multiplied by $0.6 \pm 0.2$ to estimate $T_{90}$.  In three cases, the
25 keV--1 MeV fluence was calculated from the reported 25--100 keV
fluence assuming a spectral slope of $-1$.  Squares with large error bars
show the average fluences and durations obtained by Mukherjee \etal (1998) for
203 short GRBs and 486 long GRBs, respectively. 
The data indicate that the fluence
of long bursts with observed afterglows is on average $\sim 40$ times
larger than the fluence of short GRBs. }
\end{figure*}

\begin{figure*}
\centerline{\psfig{figure=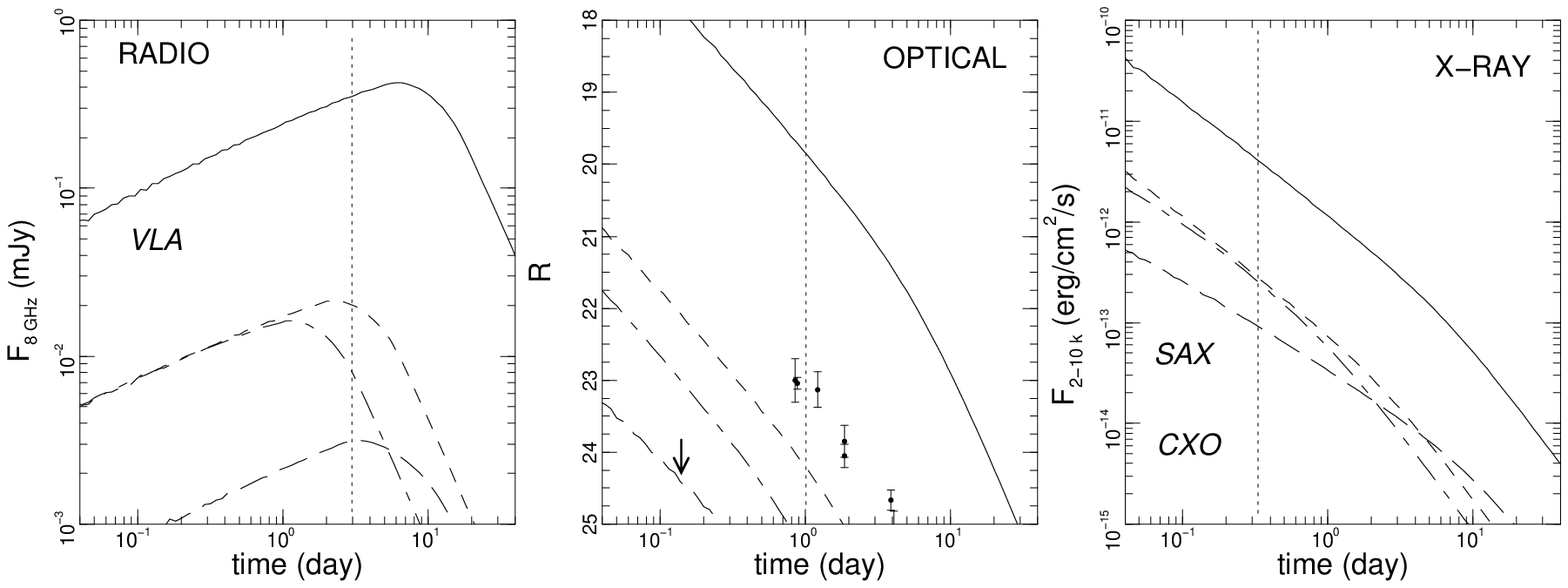}} \figcaption{The solid lines show
the radio (8 GHz), optical (R-band) and X-ray fluxes (2-10 keV band)
corresponding to a model afterglow of a long GRB.  Model parameters
are (see text): $n =0.1\, \cm3$, $\epsel = 0.03$, $\epsmag = 10^{-3}$,
$p = 2.0$, redshift $z=1$, isotropic equivalent kinetic energy $E =
10^{53}$ erg.  The relativistic ejecta were assumed to be collimated,
with an initial aperture (half-angle) of 0.1 rad.  The collimation
causes the breaks seen in some of the light-curves.  The short-dashed
lines show the model light-curves of a short GRB, for which all the
parameters are the same except that $E = 5\times 10^{51}$ erg.  The
long-dashed lines correspond to another short-GRB model in which $n =
10^{-3}\, \cm3$, more characteristic of the low density environment in
which these bursts may occur if they are associated with binary
mergers.  The dot-dashed lines are for a short GRB with a lower
redshift ($z=0.5$) and correspondingly less energy ($E = 10^{51}$
erg), as suggested by the higher value of $<V/V_{max}>$ for short GRBs
estimated by Katz \& Canel (1996), Piran (1996) and Tavani (1998).  
Also indicated on the plots are the sensitivities of the best radio and X-ray
instruments currently available, the optical upper limit (arrow)
obtained by Groot \etal (1997) for the afterglow of GRB 970828 ($R >
23.8$ at 4 hours), and the light-curve of the very dim afterglow of
the long GRB 000630 (Fynbo \etal 2001) .  The vertical dotted lines
indicate the time after the initial GRB when afterglow observations
are typically made. }
\end{figure*}


\begin{references}

\reference{} Eichler, D. \etal 1989, Nature, 340, 126
\reference{} Fryer, C., Woosley, S., Herant, M \& Davies, M. 1999, ApJ, 520, 650
\reference{} Fynbo, J. \etal 2001, A\&A 369, 373
\reference{} Goodman, J. 1986, ApJ, 308, L17
\reference{} Groot, P. \etal 1998, ApJ, 493, L27
\reference{} Hurley, K. \etal 2001, ApJ, submitted (astro-ph/0107188)
\reference{} Katz, J. 1994, ApJ, 422, 248
\reference{} Katz, J. \& Canel, L. 1996, ApJ, 471, 915 
\reference{} Kouveliotou, C. \etal 1993, ApJ, 413, L101
\reference{} Kumar, P. 1999, ApJ, 523, L113
\reference{} Kumar, P. \& Piran, T. 2000, ApJ, 535, 152
\reference{} Kumar, P. \& Panaitescu, A. 2000, ApJ, 541, L9
\reference{} Lazzati, D., Covino, S. \& Ghisellini, G. 2001, A\&A, submitted (astro-ph/0011443)
\reference{} Lee, T. \& Petrosian, V. 1997, ApJ, 474, 37
\reference{} MacFadyen, A. \& Woosley, S. 1999, ApJ, 524, 262
\reference{} MacFadyen, A., Woosley, S. \& Heger, A. 2001, ApJ, 550, 410
\reference{} Mao, S., Narayan, R., and Piran, T., 1994, ApJ, 420, 171
\reference{} Meegan, C. \etal 1996, ApJ, S106
\reference{} M\'esz\'aros, P. \& Rees, M. 1992, ApJ, 397, 570
\reference{} M\'esz\'aros, P., Rees, M.J. \& Wijers, R. 1999, New Ast, 4, 303
\reference{} Mukherjee, S. \etal 1998, ApJ, 508, 314
\reference{} Narayan, R., Paczy\'nski, B. \& Piran, T. 1992, ApJ, 395, L83
\reference{} Narayan, R., Piran, T. \& Kumar, P. 2001, ApJ, in press (astro-ph/0103360) 
\reference{} Paczy\'nski, B. 1991, Acta Ast, 41, 257 
\reference{} Paczy\'nski, B. 1998, ApJ, 494, L45
\reference{} Panaitescu, A. \& Kumar, P. 2000, ApJ, 543, 66
\reference{} Panaitescu, A. \& Kumar, P. 2001a, ApJ, 554, 667
\reference{} Panaitescu, A. \& Kumar, P. 2001b, ApJ, submitted
\reference{} Piran, T. 1996, in "Compact stars in binaries", proceedings from 
       IAU symposium 165, Eds J. van Paradijs, E. P. J. van den Heuvel, E.
                    Kuulkers. Kluwer Academic Publishers, Dordrecht, p.489 
\reference{} Piran, T. 1999, Phys. Rep., 314, 575
\reference{} Piran, T., Shemi, A. \& Narayan, R. 1993, MNRAS, 263, 861
\reference{} Ramirez-Ruiz, E. \& Fenimore, E. 2000, ApJ, 539, 712
\reference{} Rees, M.J. \& M\'esz\'aros, P. 1992, MNRAS, 258, 41p
\reference{} Rhoads, J. 1999, ApJ, 525, 737
\reference{} Sari, R., Piran, T. 1997, ApJ, 485, 270
\reference{} Sari, R., Piran, T., \& Halpern, J. 1999, ApJ, 519, L17
\reference{} Sari, R., Piran, T., \& Narayan, R. 1998, ApJ, 497, L17
\reference{} Tavani, M. 1998, ApJ, 497, L21
\reference{} Woosley, S. 1993, ApJ, 405, 273
\reference{} Zhang, W. \& Fryer, C. 2001, ApJ, 550, 357

\end{references}
\end{document}